\documentclass[12pt]{article}
\pdfoutput=1
\usepackage{jheppub}
\usepackage{mathtools,amssymb,amsthm}
\usepackage[utf8]{inputenc}
\usepackage{graphics}

\def\be{\begin{equation}}
\def\ee{\end{equation}}
\def\ba#1\ea{\begin{align}#1\end{align}}
\def\bg#1\eg{\begin{gather}#1\end{gather}}
\def\bm#1\em{\begin{multline}#1\end{multline}}
\def\bmd#1\emd{\begin{multlined}#1\end{multlined}}

\def\a{\alpha}

\def\d{\delta}

\def\e{\epsilon}

\def\g{\gamma}
\def\G{\Gamma}

\def\p{\phi}

\def\r{\rho}

\def\la{\label}

\def\re{\ref}
\def\er{\eqref}

\def\fr{\frac}

\def\td{\tilde}

\def\eq{\equiv}

\def\cd{\cdots}
\def\ap{\approx}
\def\nn{\nonumber}
\def\qu{\quad}
\def\qqu{\qquad}

\def\({\left(}
\def\){\right)}
\def\[{\left[}
\def\]{\right]}
\def\<{\langle}
\def\>{\rangle}

\def\bea{\begin{eqnarray}}
\def\eea{\end{eqnarray}}

\newcommand{\Tr}{\operatorname{Tr}}

\newcommand{\cN}{{\mathcal N}}
\newcommand{\cO}{{\mathcal O}}

\def\RR{{\mathbb R}}

\def\CC{{\mathbb C}}

\def\cF{{\cal F}}
\def\cS{{\cal S}}
\def\Li{{\rm Li}}

\def\ep{{\varepsilon}}
\def\p{\partial}

\def\sm{\smallskip}
\def\Re{{\rm Re}}

\begin{document}

\title{An Alternative Method for Extracting the \\ von Neumann Entropy from R\'enyi Entropies}
\author{Eric D'Hoker$^{(a)}$,}
\affiliation[(a)]{Mani L. Bhaumik Institute for Theoretical Physics, Department of Physics and Astronomy,\\
University of California, Los Angeles, CA 90095, USA}
\emailAdd{dhoker@physics.ucla.edu}
\author{Xi Dong${}^{(b)}$, and Chih-Hung Wu${}^{(b)}$}
\affiliation[(b)]{Department of Physics, University of California, Santa Barbara, CA 93106, USA}
\emailAdd{xidong@ucsb.edu}
\emailAdd{chih-hungwu@physics.ucsb.edu}

\abstract{An alternative method is presented for extracting the von Neumann entropy $-\operatorname{Tr} (\rho \ln \rho)$ from $\operatorname{Tr} (\rho^n)$ for integer $n$ in a quantum system with density matrix~$\rho$. Instead of relying on direct analytic continuation in $n$, the method uses a generating function $-\operatorname{Tr} \{ \rho \ln [(1-z \rho) / (1-z)] \}$ of an auxiliary complex variable $z$. The generating function has a Taylor series that is absolutely convergent within $|z|<1$, and may be analytically continued in $z$ to $z = -\infty$ where it gives the von Neumann entropy. As an example, we use the method to calculate analytically the CFT entanglement entropy of two intervals in the small cross ratio limit, reproducing a result that Calabrese et~al.\ obtained by direct analytic continuation in $n$. Further examples are provided by numerical calculations of the entanglement entropy of two intervals for general cross ratios, and of one interval at finite temperature and finite interval length.}

\maketitle

%%%%%%%%%%%%%%%%%%%%%%%%%%%%%%%%%%%%%%%%%%%%%%%%%%
\section{Introduction}\la{intro}

The von Neumann entropy provides a quantitative measure of entanglement in a quantum system.  In the special case of a thermal ensemble specified by a Hamiltonian, the von Neumann entropy reduces to the standard statistical mechanics entropy.  In quantum field theory, the thermal entropy may be obtained directly from the partition function computed by standard functional integral methods, whereas the methods for calculating the entanglement entropy of a general subsystem are much less systematic, even in free field theories.

\sm

Quantum field theory does provide, however, a reasonably systematic method for evaluating the R\'enyi entropy~\cite{Renyi:1961}
\bea
S_\nu (\rho) = \frac{1}{1-\nu} \ln \Tr (\rho^\nu)
\eea
when $\nu$ is an integer $n$ greater than $1$, by replicating the functional integral representation for $\r$, the density matrix, $n$ times.  The von Neumann entropy
\bea
S(\rho) = - \Tr (\rho \ln \rho)
\eea
is then obtained by taking the $\nu \to 1$ limit of the R\'enyi entropy:
\be
S(\rho) = \lim _{\nu \to 1} S_\nu (\rho).
\ee
Before taking such a limit, note that even though the R\'enyi entropy $S_\nu$ is well-defined for both integer and non-integer $\nu$, it is directly computed by replicating the functional integrals only when $\nu$ is an integer $n>1$.  Therefore, one needs to determine the full function $S_\nu$ from the set of its integer values $\{S_n, n=2, 3, \cdots\}$, via some kind of analytic continuation.  In practice, one achieves this analytic continuation by finding a locally holomorphic function $\cS_\nu$ of $\nu \in \CC$ that takes the values $\cS_n = S_n (\rho)$ for all integers $n>1$ and has suitable asymptotic behaviors as $\nu \to \infty$ as required by Carlson's theorem\footnote{Carlson's theorem states that a function $f(\nu)$ that is holomorphic in $\nu$ for $\Re(\nu)\geq 1$, vanishes for positive integer $\nu$, is bounded by $|f(\nu)|< C\, e^{\lambda |\nu|}$ for $\Re(\nu) \geq 1$ with constant $C, \lambda$, and satisfies this bound with some $\lambda <\pi$ on $\Re(\nu)=1$, must vanish identically for all $\nu$. See~\cite{Boas1954, Witten_2019}.}.  If such a function $\cS_\nu$ is found, Carlson's theorem guarantees its uniqueness and thus we obtain $S_\nu(\rho) = \cS_\nu$.

This replica method has been applied extensively to the calculation of the entanglement entropy in many quantum systems, including quantum field theories and conformal field theories (CFTs). For example, the entanglement entropy of a single interval of length $L$ in the vacuum state of a two-dimensional CFT on an infinite line is given by the universal formula
\bea
S(\rho)= { c \over 3} \, \ln \left ( { L \over \ep} \right ) ,
\eea
where $c$ is the central charge and $\ep$ is a UV cutoff with dimension of length. 

\sm

In this paper, we present an alternative method for extracting the von Neumann entropy from the R\'enyi entropies $S_n (\rho)$ for integer $n>1$.  Our method does not rely on a direct analytic continuation in the variable $n$.  Instead, the starting point is that once we know the traces of powers of the density matrix
\be
R_n(\rho) \eq \Tr (\rho^{n}),
\ee
we are allowed to define a generating function of an auxiliary variable $z$ for these $R_n(\rho)$:
\bea\la{gdf}
G(z;\rho) \eq - \Tr \left ( \rho \, \ln { 1-z \rho \over 1-z} \right )
= \sum _{n=1}^\infty { z^n \over n} \Big ( \Tr (\rho^{n+1} ) -1 \Big ).
\eea
For a density matrix $\rho$, the series is absolutely convergent in the unit disc $|z|<1$. Choosing the branch cut of the logarithm to lie along the positive real axis, the function $G(z;\rho)$ may be analytically continued from the unit disc to a holomorphic function in the cut plane $\CC \setminus [1, \infty)$. The  limit of this analytically continued function as $z \to - \infty$, which is well within the domain of holomorphicity, gives the von Neumann entropy:
\bea
S(\rho) = \lim _{z \to -\infty} G(z;\rho),
\eea
as may be verified directly from the definition~\er{gdf}.  The existence of the analytically continued function in $z$ beyond the unit disc may be seen from~\er{gdf} as well.  It may also be verified by rewriting the generating function in terms of a M\"obius transformed variable $w$:
\bea
G(z;\rho)=  - \Tr \Big ( \rho \ln \big \{ 1-w(1-\rho) \big \}  \Big ), \qqu
w = { z \over z-1}.
\eea
Its Taylor series in powers of $w$ is
\be
G(z;\rho) = \sum_{k=1}^\infty \fr{\td f(k)}{k} w^k
\ee
where 
\be 
\td f(k) = \Tr [\r (1-\r)^k] = \sum_{m=0}^k \fr{(-1)^m k!}{m! (k-m)!} \Tr (\r^{m+1}).
\ee
The first few terms of the series written in terms of $R_n = \Tr (\rho^{n})$ are
\be
G(z;\rho) = (1-R_2) w + \fr{1}{2} (1-2 R_2+R_3) w^2 + \fr{1}{3} (1-3 R_2 +3 R_3 -R_4) w^3 + \cd.
\ee
For a density matrix $\rho$, the Taylor series is absolutely convergent in the unit disc $|w|<1$.  Transforming back to the $z$ variable, this provides explicitly the analytic continued function in $\Re(z)\leq 1/2$.  An advantage of working with the $w$ variable is that we may obtain the von Neumann entropy directly by taking the $w \to 1$ limit:
\be
S(\r) = \lim _{w \to 1} G(z;\rho).
\ee
This may be written explicitly as a series:
\be\la{sft}
S(\r) = \sum_{k=1}^\infty \fr{\td f(k)}{k} = (1-R_2) + \fr{1}{2} (1-2 R_2+R_3) + \fr{1}{3} (1-3 R_2 +3 R_3 -R_4) + \cd
\ee
which provides an exact expression for evaluating the von Neumann entropy from $R_n = \Tr (\rho^{n})$ with integer $n>1$. 

In addition to extracting the von Neumann entropy, it is worth noting that $R_\nu (\rho) = \Tr ( \rho^\nu)$ for arbitrary powers $\nu \in \RR^+$ may be evaluated by using similar generating functions as well, whenever the corresponding trace is convergent. 

Our method is related to the resolvent method used in e.g.~\cite{Penington:2019kki}.  One advantage of our method is that it can be applied directly to numerical calculations as we will demonstrate using Eq.~\er{sft} shortly.

\sm

The remainder of this paper is organized as follows. In section~\ref{sec1a} we follow the procedure described above to recover the von Neumann entropy in the simple case of a single interval, whose solution by analytic continuation in $n$ is immediate. In section~\ref{sec2}, we use our method to evaluate the von Neumann entropy of two intervals in the small cross ratio limit in 2d CFT starting from $R_n (\rho) = \Tr ( \rho^n)$, and reproduce the results of~\cite{Calabrese:2010he} for this system in the same limit. In section~\ref{sec3}, we set up the basis for numerical calculations of the von Neumann entropy using the generating function. In section~\ref{sec4}, we present numerical results for the entanglement entropy in the two-interval example but now for finite values of the cross ratio. In section~\ref{sec5}, we present numerical results for the entanglement entropy of one interval in a two-dimensional CFT at finite temperature. In section~\ref{sec6}, we analyze in detail the rate of convergence of the Taylor series for the generating function in the M\"obius transformed variable $w$ at $w=1$.  We conclude and comment on a few open questions in section~\ref{secdis}.

%%%%%%%%%%%%%%%%%%%%%%%%%%%%%%%%%%%%%%%%%%%%%%%%%%
\section{Analytical calculation for one interval} \la{sec1a}

We now apply the method to our first example: to extract the von Neumann entropy of one interval of length $L$ in the vacuum state of a two-dimensional CFT on an infinite line, from its R\'enyi entropies
\be\la{snoi}
S_n (\r) = \fr{c}{6} \(1+\fr{1}{n}\) \ln \(\fr{L}{\e}\)
\ee
for integer $n>1$.  Here $c$ is the central charge and $\ep$ is a UV cutoff.  This is an example where direct analytic continuation in $n$ is obvious, but it is nonetheless interesting to see how our method works here.  This example also serves as a starting point for more complicated examples such as the one to be analyzed in the following section.

To apply our method, we start by rewriting Eq.~\er{snoi} as
\be
R_n(\r) = \Tr (\rho^{n}) = \(\fr{L}{\e}\)^{\fr{c}{6} \(\fr{1}{n}-n\)} = e^{\(\fr{1}{n}-n\) y},
\ee
where for convenience we have defined
\be
y = \fr{c}{6} \ln \fr{L}{\e}.
\ee
We use these $R_n(\r)$ values for integer $n$ to form the generating function
\be\la{goi}
G(z;\rho) = \sum_{n=1}^\infty {z^n \over n} \[ e^{\(\fr{1}{n+1}-n-1\)y} -1 \].
\ee
Combining $e^{-ny}$ with $z^n$ and expanding the remaining exponential in powers of $y$, we obtain
\ba
G(z;\rho) &= \ln(1-z) + e^{-y} \sum_{n=1}^\infty {(z e^{-y})^n \over n} \sum_{j=0}^\infty \fr{y^j}{j!} \fr{1}{(n+1)^j}
 \\
&= \ln(1-z) + e^{-y} \sum_{j=0}^\infty \fr{y^j}{j!} F_j(z e^{-y}), \la{goie}
\ea
where we have defined the sum
\be\la{snp}
F_j(z) \eq \sum_{n=1}^\infty \fr{z^n}{n (n+1)^{j}}
\ee
for nonnegative integer $j$.
We will eventually take $z\to -\infty$, so we need to find the behavior of the sum~\er{snp} in this limit.  To do this, we use the partial fraction decomposition
\be
\label{parfrac}
{1 \over n (n+1)^{k}} = {1 \over n} - \sum _{s=1}^{k} {1 \over (n+1)^s}
\ee
and perform the sum~\er{snp} in terms of the polylogarithm function defined by 
\bea
\label{Lik}
\Li_s(z) = \sum _{n=1}^\infty { z^n \over n^s},
\eea
obtaining
\be\la{fkp}
F_j(z) = j + \Li_1(z) -\fr{1}{z} \sum_{s=1}^j \Li_s(z).
\ee

To find the behavior of $F_j(z)$ as $z \to - \infty$, we need the behavior of the polylogarithm as $z \to -\infty$, which is given by the Sommerfeld expansion
\bea
\Li_s(z) = - 2 \sum_{m=0}^\infty  
{  \(1 - 2^{1-2m}\) \zeta(2m) \,  \over \Gamma (s-2m+1)} \[\ln(-z)\]^{s-2m} + \cO(z^{-1}).
\eea
From this we find
\be\la{fkl}
F_j(z) = j - \ln(-z) + \cO(z^{-1})
\ee
as $z \to -\infty$.  Substituting this into Eq.~\er{goie}, we obtain as $z \to -\infty$
\ba
G(z;\rho) &= \ln(1-z) + e^{-y} \sum_{j=0}^\infty \fr{y^j}{j!} \[j - \ln(-ze^{-y}) + \cO(z^{-1})\]\\
&= \ln(-z) + y - \ln(-ze^{-y}) + \cO(z^{-1})\\
&= 2y + \cO(z^{-1}).
\ea
Taking the $z \to - \infty$ limit, we obtain the von Neumann entropy
\be
S(\r) = \lim_{z\to-\infty} G(z;\rho) = 2y = \fr{c}{3} \ln \fr{L}{\e}
\ee
as expected.

%%%%%%%%%%%%%%%%%%%%%%%%%%%%%%%%%%%%%%%%%%%%%%%%%%
\section{Analytical calculation for two intervals} \la{sec2}

We now apply our method to calculate the entanglement entropy of two disjoint intervals in the vacuum state of a two-dimensional CFT on an infinite line.  Let us call the two-interval subsystem $A\cup B$, where $A=[x_1, x_2]$ and $B=[x_3, x_4]$.  The cross ratio is then defined as
\be
x=\frac{x_{12}x_{34}}{x_{13}x_{24}},
\ee
with $x_{ij}=x_{i}-x_{j}$.  We will focus on the small $x$ limit, where $R_n (\rho) = \Tr (\rho^n)$ was calculated in~\cite{Calabrese:2009ez, Calabrese:2010he} and given by
\be\la{trn2}
\Tr (\rho^n) = \(\fr{x_{12} x_{34}}{\e^2}\)^{\fr{c}{6}\(\fr{1}{n}-n\)} \[ 1+\mathcal{N} \left ( { x \over 4 n^2} \right )^\alpha \fr{n}{2} \sum_{\ell=1}^{n-1} {1 \over \left ( \sin {\pi \ell \over n} \right )^{2 \alpha}} + \cd \],
\ee
where $\e$ is a UV cutoff, $\a/2$ is the lowest dimension in the operator spectrum of the CFT, $\cN$ is the multiplicity of the lowest-dimensional operators, and $\cd$ denotes higher order corrections.  For examples, we have $\cN=2$ for a free boson and $\cN=1$ for the Ising model.   For notational simplicity, let us define
\be
y = \fr{c}{6} \ln \fr{x_{12} x_{34}}{\e^2}
\ee
and write the sum in Eq.~\er{trn2} in a different but equivalent way, leading to
\be\la{trn2b}
\Tr (\rho^n) = e^{\(\fr{1}{n}-n\)y} \[ 1+\mathcal{N} \left ( { x \over 4 n^2} \right )^\alpha \sum_{\ell=1}^{n-1} {\ell \over \left ( \sin {\pi \ell \over n} \right )^{2 \alpha}} +\cd \].
\ee

At the leading order $\cO(x^0)$, the form of $\Tr (\r^n)$ is similar to that of a single interval studied in section~\re{sec1a}, and therefore the von Neumann entropy may be extracted by the same steps, resulting in
\be\la{slead}
S_{AB} = 2 y + \cO(x^\a) = \fr{c}{3} \ln \fr{x_{12} x_{34}}{\e^2} + \cO(x^\a).
\ee
We now work at the subleading order $\cO(x^\a)$.  As our method deals with $\Tr (\r^n)$ in a completely linear way, we may focus on the $\cO(x^\a)$ term in Eq.~\er{trn2b}.  To extract the von Neumann entropy at this order, we define the generating function at order $x^\a$
\bea
\label{tiG1}
\tilde G(z;\rho) = \sum _{n=1}^ \infty { z^n \over n} e^{\(\fr{1}{n+1}-n-1\)y} { 1 \over (n+1)^{2 \alpha}} \sum_{\ell=1}^n { \ell \over \left ( \sin {\pi \ell \over n+1} \right )^{2 \alpha}},
\eea
where we have removed a multiplicative factor $\cN \(\fr{x}{4}\)^\a$ for notational simplicity\footnote{We may also strip off the ``easy factor'' $e^{\(\fr{1}{n+1}-n-1\)y}$  from the generating function~\er{tiG1}.  This factor is easy to analytically continue in $n$ and was separately treated in section~\ref{sec1a}.  It is not completely obvious that stripping it off would lead to the correct final answer for the von Neumann entropy, but it actually does (as one may check as in footnote~\re{ft2}), suggesting that our method has a broader regime of applicability than what might be expected from its derivation in section~\re{intro}.\la{ft1}}.

Our aim is to evaluate this function and analytically continue it to $z =-\infty$. To this end we begin by deriving an integral representation.

\subsection{Integral representation for $\tilde G$}

We first combine $e^{-ny}$ with $z^n$ in Eq.~\er{tiG1} and expand the remaining exponential in powers of $y$ as in section~\re{sec1a}, obtaining
\ba
\tilde G(z;\rho) &= e^{-y} \sum _{n=1}^ \infty {(z e^{-y})^n \over n} \sum_{j=0}^\infty \fr{y^j}{j!} { 1 \over (n+1)^{2 \alpha+j}} \sum_{\ell=1}^n {\ell \over \left ( \sin {\pi \ell \over n+1} \right )^{2 \alpha}}.
\ea
We now use the expansion proposed in~\cite{Calabrese:2009ez} for the functions
\bea
\label{pk}
\left ( { u \over \sin u} \right )^{2 \alpha} = \sum _{k=0}^\infty p_k(\alpha) \, u^{2k}
\eea
where $p_k(\alpha)$ is a polynomial in $\alpha$ of degree $k$. The radius of convergence of this Taylor expansion in $u$ is $\pi$, where the first singularity away from $u=0$ is located. Substituting $u =  \pi \ell /(n+1)$, we obtain the following sum
\be
\label{tiG2}
\tilde G(z;\rho) =
\fr{e^{-y}}{\pi} \sum _{k=0}^\infty p_k (\alpha) \sum_{j=0}^\infty \fr{y^j}{j!} \sum _{n=1}^ \infty {(z e^{-y})^n \over n (n+1)^{j+2k}} \sum_{\ell=1}^n { 1 \over (\pi \ell)^{2\alpha - 2k-1}}.
\ee
Using a representation for the factor $(\pi \ell)^{-2 \alpha+1}$ in terms of an integral over an auxiliary variable $t$, which is convergent for all values $\Re (\alpha) >\frac{1}{2}$, and obtaining the factor $(\pi \ell)^{2k}$ by applying a derivative in $t$ of order $2k$, we have 
\bea
\sum_{\ell=1}^n { 1 \over (\pi \ell)^{2\alpha - 2k-1}} 
=
{ 1 \over \Gamma (2\alpha-1)} \int _0^\infty { d t \over t} \, t^{2 \alpha-1} \, 
\left ( { \p \over \p t} \right )^{2k} \sum _{\ell=1}^n e^{- \pi t \ell}.
\eea
Carrying out the finite geometric sum over $\ell$, and substituting the result into the expression (\ref{tiG2}), we get the following integral representation for $\tilde G(z;\rho)$:
\ba\la{gint}
\tilde G(z; \rho) &= \frac{1}{\pi}
\sum _{k=0}^\infty { p_k (\alpha) \over \Gamma (2 \alpha-1)} 
\int _0^\infty {d t \over t} \, t^{2 \alpha-1} \left ( { \p \over \p t} \right )^{2k}  \left ( { H_k(z,t) \over e^{ \pi t} -1} \right ),\\\la{hdef}
H_k (z,t) &= e^{-y} \sum_{j=0}^\infty \fr{y^j}{j!} \sum _{n=1}^ \infty {(z e^{-y})^n \over n (n+1)^{j+2k}} (1 - e^{-\pi t n}).
\ea
One verifies that the above integral representation converges absolutely for $\Re(\alpha)>\frac{1}{2}$. Indeed, for fixed $|z|<1$ and large $t$, the function $H_k(z,t)$ and all its derivatives  tend to a finite limit, so that exponential convergence of the integral in Eq.~\er{gint} is assured as $t \to \infty$. Furthermore, the functions $H_k(z,t)$ vanish linearly at $t=0$, so that the ratio $H_k(z,t)/(e^{\pi t}-1)$ and derivatives of the ratio inside the integral in Eq.~\er{gint} are integrable at $t=0$.  

We may rewrite Eq.~\er{hdef} conveniently using $F_j(z)$ defined by Eq.~\er{snp}:
\be\label{geq}
H_k (z,t) = e^{-y} \sum_{j=0}^\infty \fr{y^j}{j!} \[F_{j+2k}(ze^{-y}) - F_{j+2k}\(z e^{-y-\pi t}\)\].
\ee
Again $F_{j+2k}(z)$ may be expressed in terms of the polylogarithm according to Eq.~\er{fkp}.

\subsection{Analytic continuation to $z \to - \infty$}

To compute the limit of $\tilde G(z; \rho)$ as $z \to - \infty$, we need the behavior of $F_{j+2k}(z)$ as $z \to -\infty$, which is given by Eq.~\er{fkl}:
\be
F_{j+2k}(z) = j+2k - \ln(-z) + \cO(z^{-1}).
\ee
In this limit, Eq.~\er{geq} becomes\footnote{The dependence on $y$ drops out in Eq.~\er{hks}, leaving us with what we would have obtained if we had stripped off the ``easy factor'' $e^{\(\fr{1}{n+1}-n-1\)y}$ from the generating function~\er{tiG1}, thus confirming our claim in footnote~\re{ft1}.\la{ft2}}
\ba
H_k(z,t) &= e^{-y} \sum_{j=0}^\infty \fr{y^j}{j!} \[-\ln(ze^{-y}) + \ln(z e^{-y-\pi t}) + \cO(z^{-1})\]\\\la{hks}
&= -\pi t + \cO(z^{-1}).
\ea
As a result, the evaluation of $\tilde G(z; \rho)$ in this limit reduces to
\bea
\label{tiG3}
\lim _{z \to - \infty} \tilde G(z; \rho) =
-  \sum _{k=0}^\infty  p_k (\alpha) g_k (\alpha) 
\eea
where the coefficients $g_k(\alpha) $ are given by the following integral representation
\bea
g_k (\alpha) =  { 1 \over \Gamma (2 \alpha-1)} \int _0^\infty {d t \over t} \, t^{2 \alpha-1} \left ( { \p \over \p t} \right )^{2k}  \left ( { t \over e^{ \pi t} -1} \right ).
\eea
The integral is absolutely convergent for all $\Re(\alpha ) >\frac{1}{2}$. To evaluate it, we shall use analytic continuation in $\alpha $ to complex values  $\Re (\alpha ) >k+\frac{1}{2}$ such that we may integrate by parts $2k$ times (with vanishing boundary terms), and obtain the following expression
\bea
g_k (\alpha) =  \pi^{2k-2\alpha } (2 \alpha - 2k-1) \zeta(2 \alpha -2k).
\eea
For fixed $\alpha$ and $k+\frac{1}{2} > \Re (\alpha)$, the sign of $g_k(\alpha)$ alternates as a function of $k$ and its magnitude grows with $k$. Thus the sum (\ref{tiG3}) is an asymptotic expansion with alternating coefficients, which we will evaluate in the next subsection using a procedure similar to Borel resummation.

\subsection{Matching with previous results}

To make contact with the results of Calabrese, Cardy, and Tonni in~\cite{Calabrese:2010he}, we begin by using the functional relation of the Riemann $\zeta$-function
\bea
\zeta (1-s) = { 2 \Gamma (s) \over (2 \pi)^s} \cos {\pi s \over 2} \, \zeta (s)
\eea
to express $\zeta (2 \alpha - 2k ) $ in terms of $\zeta (2k-2\alpha+1)$, and then use the standard integral representation 
\bea
\Gamma (s) \zeta (s) = \int _0^\infty { dt \over t} \, { t^s \over e^t -1}
\eea
for $\zeta (2k-2\alpha+1)$, to recast $g_k (\alpha)$ as follows:
\bea
g_k (\alpha) = \frac{2}{\pi} (-)^{k+1} \sin ( \pi \alpha) \int _0^\infty  { dt \over e^{2t} -1} \, { \p \over \p t} \, t^{2k-2\alpha+1}.
\eea 
Substituting this representation into (\ref{tiG3}), we find
\bea
\label{inter2}
\lim _{z \to - \infty} \tilde G(z; \rho) =
\frac{2 \sin (\pi \alpha) }{\pi}   \int _0^\infty  { dt \over e^{2t} -1} \, { \p \over \p t}  \sum _{k=0}^\infty  p_k (\alpha) (-)^k t^{2k-2\alpha+1}. 
\eea
The sum over $k$ is easily recognized in terms of the defining relation (\ref{pk}) of $p_k (\alpha)$:
\bea
\sum _{k=0}^\infty  p_k (\alpha) (-)^k t^{2k-2\alpha} =  { 1 \over (\sinh t)^{2 \alpha}}.
\eea
Substituting this into (\ref{inter2}), we find a much simplified integral representation:
\bea
\label{inter3}
\lim _{z \to - \infty} \tilde G(z; \rho) =
\frac{2 \sin (\pi \alpha) }{\pi}    \int _0^\infty  { dt \over e^{2t} -1} \, { \p \over \p t}  { t \over (\sinh t)^{2 \alpha}}.
\eea
The integral may be simplified upon integrating by parts, and we obtain
\bea
\label{inter4}
\lim _{z \to - \infty} \tilde G(z; \rho) =
\frac{\sin (\pi \alpha)}{\pi}    \int _0^\infty  dt { t \over  (\sinh t)^{2 \alpha+2}}.
\eea
The integral is absolutely convergent for $-1<\Re(\alpha)<0$ and may be evaluated exactly for this range of parameters: 
\bea
\int _0^\infty  dt { t \over  (\sinh t)^{2 \alpha+2}}=\frac{\sqrt{\pi}}{4} \cot{(\pi \alpha)} \Gamma(-\alpha-\tfrac{1}{2}) \Gamma(\alpha+1).
\eea
Applying  the reflection formula for the $\Gamma$-function 
\be
 \Gamma(-\alpha-\tfrac{1}{2})=-\frac{\pi}{\Gamma(\alpha+\frac{3}{2}) \cos(\pi \alpha)},
\ee
we see that the pre-factor $\cot (\pi \alpha)$ cancels  both the pre-factor $\sin (\pi \alpha)$ in (\ref{inter4}) and the factor  $\cos (\pi \alpha)$ from the reflection formula, and we find
\bea
\lim _{z \to - \infty} \tilde G(z; \rho) = - \frac{\sqrt{\pi} \Gamma(\alpha+1)}{4 \Gamma (\alpha+\frac{3}{2})}.
\eea
Restoring the multiplicative factor $\cN \(\fr{x}{4}\)^\a$ that we ignored near the beginning of the calculation, and combining this with the leading order result~\er{slead}, we find the von Neumann entropy of two intervals
\be
S_{AB} = \fr{c}{3} \ln \fr{x_{12} x_{34}}{\e^2} - \cN  \frac{\sqrt{\pi} \Gamma(\alpha+1)}{4 \Gamma (\alpha+\frac{3}{2})} \(\fr{x}{4}\)^\a +\cd
\ee
where $\cd$ denotes higher order corrections.  Note that the first term is simply $S_A+S_B$, so the remaining terms give $-I_{AB}$ where $I_{AB}$ is the mutual information.  This coincides exactly with the result of~\cite{Calabrese:2010he} obtained by direct analytic continuation in $n$. 

%%%%%%%%%%%%
\section{Setup of the numerical method} \la{sec3}

While the application of our method in various examples of conformal field theory is highly nontrivial, it turns out that the numerical application of our method provides promising approximations in a variety of complicated models. In the present section, we shall establish the numerical procedure for the next two sections. 

Given our generating function
\be\la{gdef}
G(z; \rho)=\sum _{k=1}^\infty { z^k \over k} \Big( \Tr (\rho^{k+1})  -1  \Big)  \eq \sum_{k=1}^\infty \fr{f(k)}{k} z^k,
\ee
where we have defined
\be
f(k)=\Tr \left ( \rho^{k+1} \right ) -1,
\ee
the goal is to find the von Neumann entropy via the limit
\be
S = \lim_{z\to-\infty} G(z; \rho) \,.
\ee

Similarly, for two disjoint regions $A$ and $B$, we are often given 
\be\la{fkmi}
f(k) \eq \fr{\Tr \r_A^{k+1} \Tr \r_B^{k+1}}{\Tr \r_{AB}^{k+1}} - 1
\ee
but are interested in the mutual information
\be \la{mui}
I_{AB} = -f'(0) = S_A +S_B -S_{A \cup B} \,.
\ee
We conjecture\footnote{We have not yet found a general proof of this conjecture, but it is certainly plausible given the broader regime of applicability noted in footnote~\re{ft1}.  Furthermore, this conjecture will be numerically verified in section~\re{deli}.} that our method applies to the mutual information as well.  In other words, we expect to be able to use the same generating function~\er{gdef} with $f(k)$ given by Eq.~\er{fkmi} and obtain the mutual information
\be
I_{AB} = \lim_{z\to-\infty} G(z; \rho) \,,
\ee
where $\r$ can be thought of as the density matrix on $A\cup B$.

The M\"obius transformation
\be\la{mobius}
w = \fr{z}{z-1} \,,\qqu
z = \fr{w}{w-1}
\ee
maps $z=1$ to $w=\infty$ and $z=\pm\infty$ to $w=1$, allowing us to rewrite the generating function as
\be\la{hw}
h(w) \eq G(z(w); \rho) = \sum_{k=1}^\infty \fr{f(k)}{k} \(\fr{w}{w-1}\)^k \,.
\ee
Supposing that $G(z; \rho)$ has singularities within $[z_1, z_2]$ with $1\leq z_1\leq z_2\leq \infty$, $h(w)$ would have singularities within $[w(z_2), w(z_1)]$ with $1\leq w(z_2)\leq w(z_1)\leq \infty$.  Therefore we have mapped a difficult problem of evaluating $G(z; \rho)$ well outside its radius of convergence (which is $z_1$) to an easy one of evaluating $h(1)$. 

The radius of convergence for $h(w)$ is $w(z_2)$, so in the worst scenario of $z_2=\infty$ we would evaluate $h(1)$ at the edge of the radius of convergence.  If $z_2$ is finite, we would evaluate $h(1)$ as an absolutely convergent series.  The value of $z_2$ is equal to the inverse of the smallest eigenvalue $\r_{\min}$ of the density matrix $\r$. Therefore we reorganize the terms in~\er{hw} as a manifest Taylor series in $w$:
\be\la{hws}
h(w) = \sum_{k=1}^\infty \fr{\td f(k)}{k} w^k \,,
\ee
where 
\be\la{tfk}
\td f(k) = \Tr [\r (1-\r)^k] = \sum_{m=0}^k \fr{(-1)^m k!}{m! (k-m)!} \Tr (\r^{m+1})
\ee
may be determined as a linear combination of $f(1)$, $f(2)$, $\cdots$, $f(k)$.

Numerically it is straightforward to evaluate~\er{hws} by truncating the Taylor series at some large order $k_{\max}$.  Therefore by knowing the R\'enyi entropies up to order $k_{\max}+1$, we can numerically estimate the von Neumann entropy with good precision.

In all the field theory examples that we will consider below, the sum~\er{hws} exhibits stable power-law behaviors at large $k$. In other words, $\td f(k)/k \ap C k^{-p}$ with some power $p>1$ at large $k$.  Intuitively, this is because in a quantum field theory, the smallest eigenvalue $\r_{\min} \to 0$ in the continuum limit and we are therefore evaluating the sum~\er{hws} at the edge of its radius of convergence, leading to power-law convergence.
We will determine the power $p$ numerically in each example below, and give an interpretation of this power in section~\ref{sec6}.

\section{Numerical studies of two intervals} \la{sec4}

Many two-dimensional CFTs may be constructed in terms of the free field theory of scalar bosons using the Coulomb gas representation and  bosonization. For those theories the basic building blocks are the CFTs of a free boson with central charge $c=1$, which are further distinguished by the compactification radius.
 In this case, the von Neumann entropies for the union of two intervals on an infinite line are characterized by the cross ratio $x$, as well as a universal critical exponent $\eta$, which is proportional to the square of the compactification radius.

For finite $x$ and $\eta$ in a free boson CFT, the general form of $\Tr(\rho^n)$ was derived in~\cite{Calabrese:2009ez}:
\be \la{ge}
\Tr(\r^n) =c_n\bigg( \frac{\epsilon^2 x_{13}x_{24}}{x_{12}x_{34}x_{14}x_{23}} \bigg)^{\frac{1}{6}(n-\frac{1}{n})} \mathcal{F}_{n}(x, \eta),
\ee
where $\e$ is a UV cutoff and $c_n$ is a non-universal, model-dependent coefficient with $c_1=1$~\cite{Calabrese:2009qy}. In our numerical calculations below, we will choose $c_n=1$ for simplicity and choose a reasonable value of $\e$.\footnote{We need to implement the cutoff and non-universal coefficient in a physical way; in particular, bounds on R\'enyi entropies such as $\Tr(\r^n) \leq 1$ should not be violated.}   Note $\mathcal{F}_{n}(x, \eta)$ is defined as
\be \la{RS}
\mathcal{F}_{n}(x, \eta)=\frac{\Theta(0 | \eta \Gamma) \Theta (0 | \Gamma/ \eta)}{[\Theta (0 | \Gamma)]^2},
\ee
for generic integers $n \geq 1$. The Riemann-Siegel theta function $\Theta(z| \Gamma)$ is defined as
\be
\Theta(z| \Gamma) \equiv \sum_{m  \in \mathbb{Z}^{n-1}} \exp[i \pi m^{t} \cdot \Gamma \cdot m+2 \pi i m^{t} \cdot z],
\ee
where $\Gamma$ is an $(n-1)\times (n-1)$ matrix with elements
\be
\Gamma_{rs}=\frac{2i}{n} \sum_{k=1}^{n-1} \sin\bigg( \frac{\pi k}{n} \bigg) \beta_{k/n} \cos \bigg[ \frac{2 \pi k}{n}(r-s) \bigg],
\ee
and
\be\la{bfdef}
\beta_{y}=\frac{F_{y}(1-x)}{F_{y}(x)},\qqu
F_{y}(x) \equiv {}_2 F_1(y,1-y;1;x),
\ee
with ${}_2 F_1$ being the hypergeometric function. Note that~\er{RS} is manifestly invariant under $\eta \leftrightarrow 1/\eta $. Currently, the analytic continuation of the von Neumann entropy for general finite $x$ and $\eta$ is not analytically known. But given our method, one can calculate the von Neumann entropy with high accuracy numerically. 

In the following two subsections, we will present the numerical studies of two intervals in the small $x$ or the decompactification $\eta \to \infty$ limit, where analytic perturbative expansions are available for comparison with our method. We will look at the general case of finite $x$, $\eta$ in the third subsection.

\subsection{Two intervals at small cross ratio}

For general $\eta \neq 1$,\footnote{For the special case of $\eta=1$, we have $\cF_n(x)=1$ instead of the small $x$ expansion~\er{ftie}.} $\mathcal{F}_{n}(x)$ has the following small $x$ expansion~\cite{Calabrese:2009ez, Calabrese:2010he}:
\be\la{ftie}
\mathcal{F}_{n}(x)=1+ \bigg( \frac{x}{4n^2} \bigg)^\alpha s_{2}(n)+\cdots \,,\qqu
s_2(n) \eq \mathcal N \fr{n}{2} \sum_{j=1}^{n-1} \fr{1}{\[\sin(\pi j/n)\]^{2\a}} \,,
\ee
where $\a$ is twice the lowest operator dimension in the CFT, $\mathcal N$ denotes its multiplicity, and $\cd$ denotes higher-order terms. For a free boson, we have $\alpha= \text{min}[\eta, 1/\eta]$ and $\mathcal N=2$. We will numerically calculate the von Neumann entropy and confirm its small $x$ expansion
\be \la{sma}
S_{A \cup B} =c'_1+ \frac{1}{3} \ln \bigg( \frac{x_{12}x_{34}x_{14}x_{23}}{\epsilon^2 x_{13}x_{24}} \bigg)-  \mathcal N  \bigg(\frac{x}{4}\bigg)^{\alpha}\fr{\sqrt{\pi} \G(\a+1)}{4\G\(\a +\fr{3}{2}\)}- \cdots,
\ee
where $c'_1$ is minus the $n$-derivative of $c_n$ at $n=1$, which is determined by matching $S_{A \cup B}$ to $S_A+S_B$ in the limit of $x_{21}, x_{43} \ll x_{31}, x_{42}$.  This small $x$ expansion of the von Neumann entropy agrees with our analytical calculation in section~\ref{sec2}.
\begin{figure} 
\centering
\includegraphics[width=0.48\textwidth]{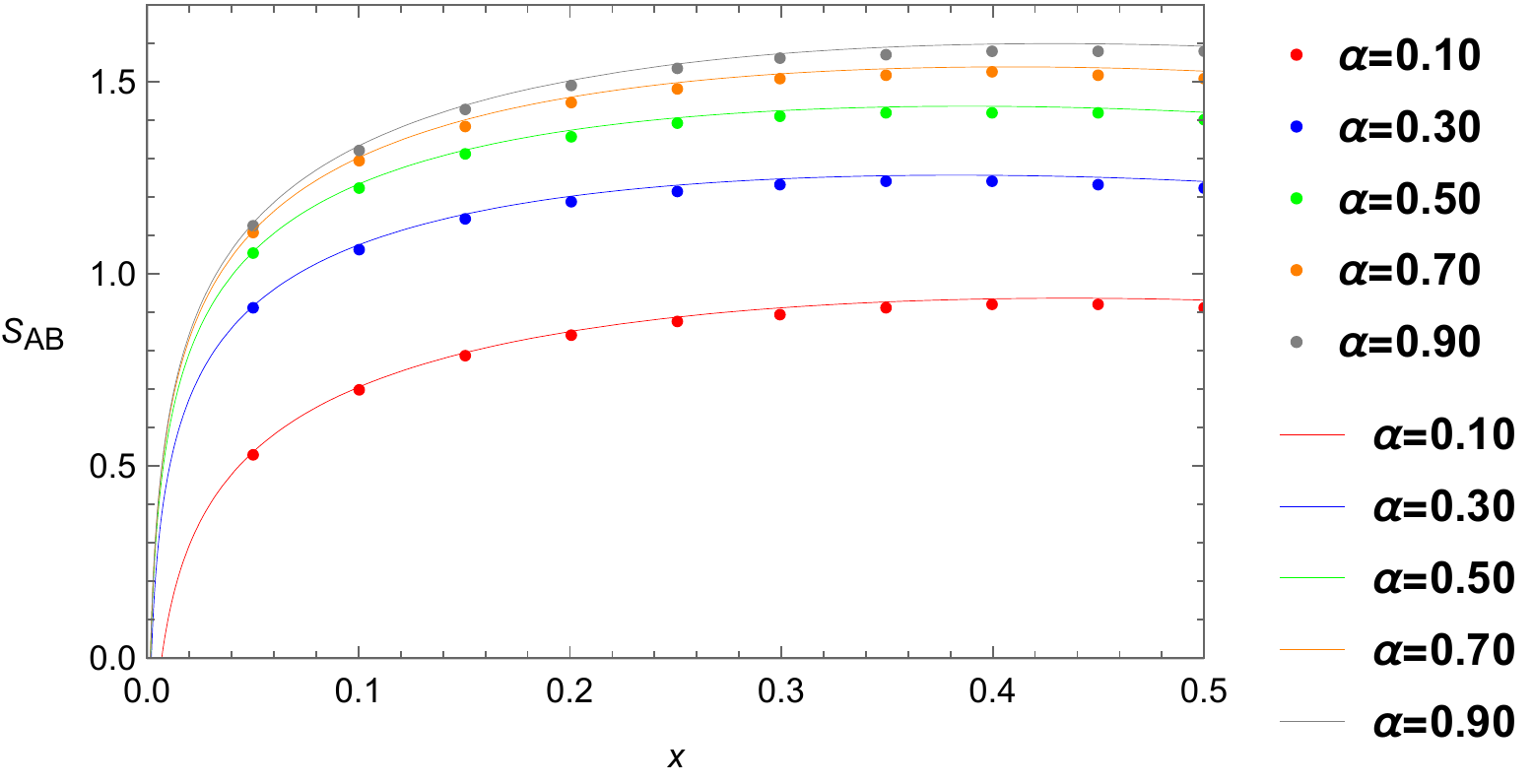}
\includegraphics[width=0.48\textwidth]{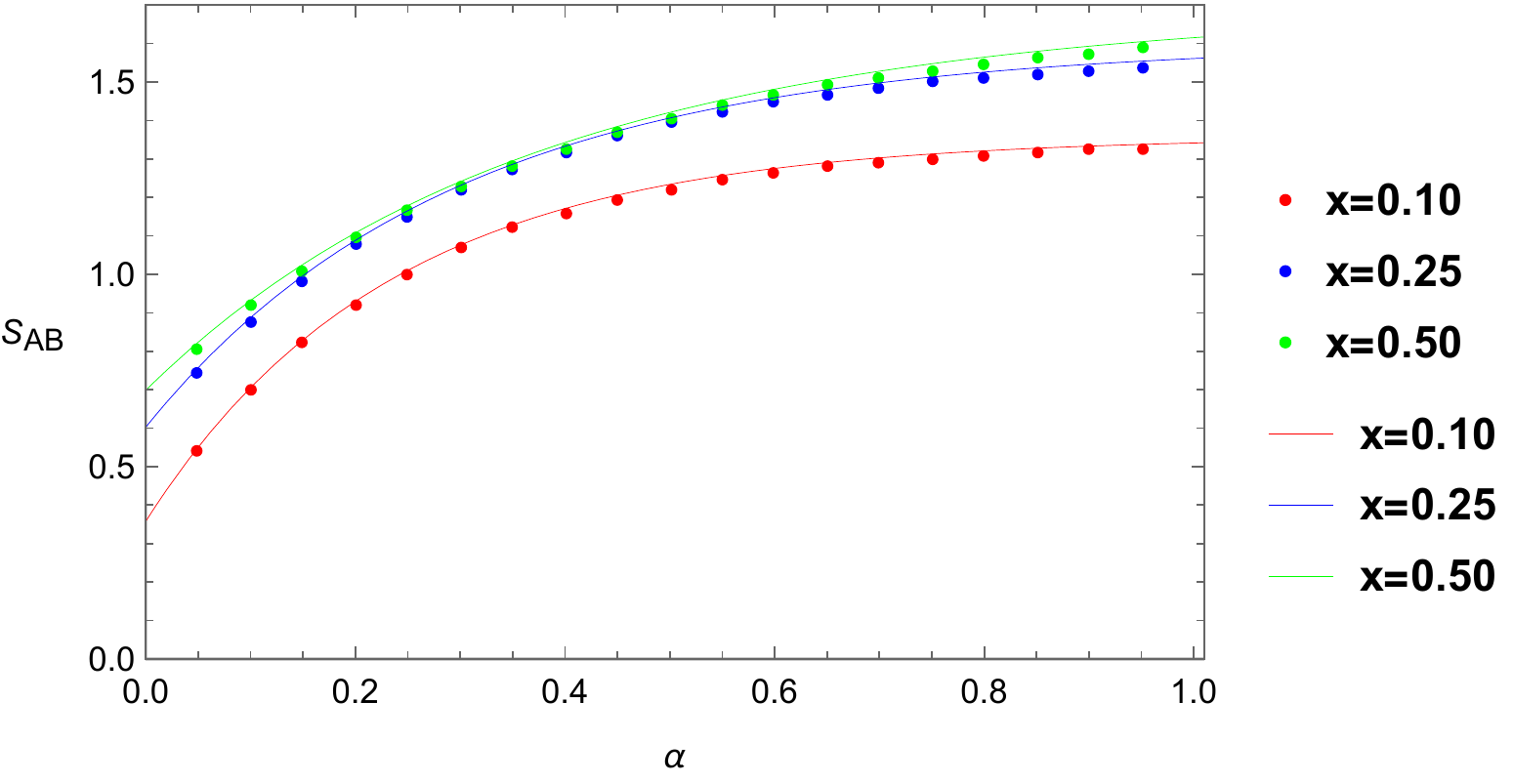}
\caption{\label{smallx} Numerical studies of the von Neumann entropy $S_{AB}$ for two intervals to first order in the small cross ratio expansion. The solid continuous curves are plotted by the analytical result~\er{sma}. The points are plotted by our numerical procedure, each of which is calculated up to $k_{\max}=400$. Left: we fix the values of $\alpha$ and plot $S_{AB}$ as function of $x$ (up to $x=0.5$). Right: we fix the values of $x$ and plot $S_{AB}$ as function of $\alpha$. We have chosen the cutoff to be $\epsilon^2=0.3$ in all cases, and one can always choose a smaller $\epsilon$ which has the effect of raising the curves.}
\end{figure}

To set up the numerics, we model the system of two disjoint intervals $A=[x_1,x_2]$ and $B=[x_3,x_4]$ on an infinite line\footnote{A similar calculation can be carried out on a circle as in the XXZ spin-chain model reported in~\cite{Furukawa:2008uk}.} with $|x_{12}|=r=|x_{34}|$, and set the distance between the centers of $A$ and $B$ to be $L=14$.  In this case, the cross-ratio is 
\be
x=\frac{x_{12}x_{34}}{x_{13}x_{24}}=\frac{r^2}{L^2}.
\ee
Note also that $|x_{14}|=L+r=L(1+\sqrt{x})$ and $|x_{23}|=L-r=L(1-\sqrt{x})$. Thus we may rewrite the ``easy factor'' in front of $\cF_n(x,\eta)$ in Eq.~\er{ge} in terms of $x$, $L$.

We have performed the numerical calculation with many choices of numerical constants. For example, with $x=0.25$, $\alpha=\eta=0.295$, and $\epsilon^2=0.3$, the result is $S_{A \cup B} \approx 1.216$ by summing up to $k_{\max}=800$, within $10^{-2}$ of the answer $S_{A \cup B} \approx 1.224$ approximated by~\er{sma}.  The sum~\er{hws} exhibits a stable power $p \approx 1.744$, with around $10^{-3}$ relative error. We have probed more regimes in the parameter space; see Figure~\ref{smallx}, where we have plotted the entanglement entropies calculated by our numerical procedure compared with the analytical result given by~\er{sma}.

Furthermore, the second order contribution in $x$ given in~\cite{Calabrese:2010he} can be numerically evaluated, even though we are not aware of an explicit analytic continuation\footnote{However, the holographic case was studied in~\cite{Barrella:2013wja,Chen:2013kpa,Chen:2013dxa}.} to $n\ap1$. In this case, we use
\be
\mathcal{F}_{n}(x)=1+ \bigg( \frac{x}{4n^2} \bigg)^\alpha s_{2}(n)+\bigg(\frac{x}{4n^2} \bigg)^{2 \alpha} s_4(n)+ \cdots,
\ee
where
\bea
s_4(n)&=&\frac{n}{2} \sum_{j_4=j_3+1}^{n-1} \sum_{j_3=j_2+1}^{n-1} \sum_{j_2=1}^{n-1} \bigg\{ \bigg[\frac{\sin{\frac{\pi (j_4 -j_2)}{n}}\sin{\frac{\pi j_3}{n}}}{\sin{\frac{\pi j_2}{n}}\sin{\frac{\pi (j_4-j_3)}{n}}\sin{\frac{\pi j_4}{n}}\sin{\frac{\pi (j_3-j_2)}{n}}} \bigg]^{2 \alpha} 
\nn\\ && \hskip 1in
+ 2 \bigg[\frac{\sin{\frac{\pi j_4 }{n}}\sin{\frac{\pi (j_3-j_2)}{n}}}{\sin{\frac{\pi j_2}{n}}\sin{\frac{\pi (j_4-j_3)}{n}}\sin{\frac{\pi (j_4-j_2)}{n}}\sin{\frac{\pi j_3}{n}}} \bigg]^{2 \alpha}   \bigg\}.
\eea
Note that the second order terms only start contributing at $n \geq 4$. Now again with $x=0.25$, $\alpha =0.295$, and $\epsilon^2=0.3$, the result is $S_{A\cup B}=1.105$ by summing up to $k_{\max}=70$.  In particular, the second order correction to the entanglement entropy is negative as suggested by the last minus sign in~\er{sma}. Since we are currently unaware of an analytical calculation at second order in $x$ for the free boson, we do not have a way to analytically confirm our numerical results in general, but our numerical study is consistent with the holographic case~\cite{Barrella:2013wja}. 

\subsection{Two intervals in the decompactification limit} \la{deli}

We now consider a different limit~\cite{Calabrese:2009ez} than the small $x$ case. In the decompactification limit $\eta \to \infty$, we have for each fixed value of $x$
\be
\mathcal{F}_{n}(x, \eta)=\bigg[ \frac{\eta^{n-1}}{\prod^{n-1}_{k=1} F_{k/n}(x) F_{k/n}(1-x)} \bigg]^\frac{1}{2},
\ee
where $F_{k/n}$ is defined as in Eq.~\er{bfdef}.  To see $\mathcal{F}_{n}(0,\eta)=1$ in the $\eta\to\infty$ limit, we note that both the numerator and the denominator go to infinity.

We will use the symmetry $\eta \leftrightarrow 1/\eta$ to study the result for $\eta \ll 1$ instead:
\be
\mathcal{F}_{n}(x, \eta)= \bigg[ \frac{\eta^{-(n-1)}}{\prod^{n-1}_{k=1} F_{k/n}(x) F_{k/n}(1-x)} \bigg]^\frac{1}{2}.
\ee

\begin{figure} 
\centering
\includegraphics[width=0.48\textwidth]{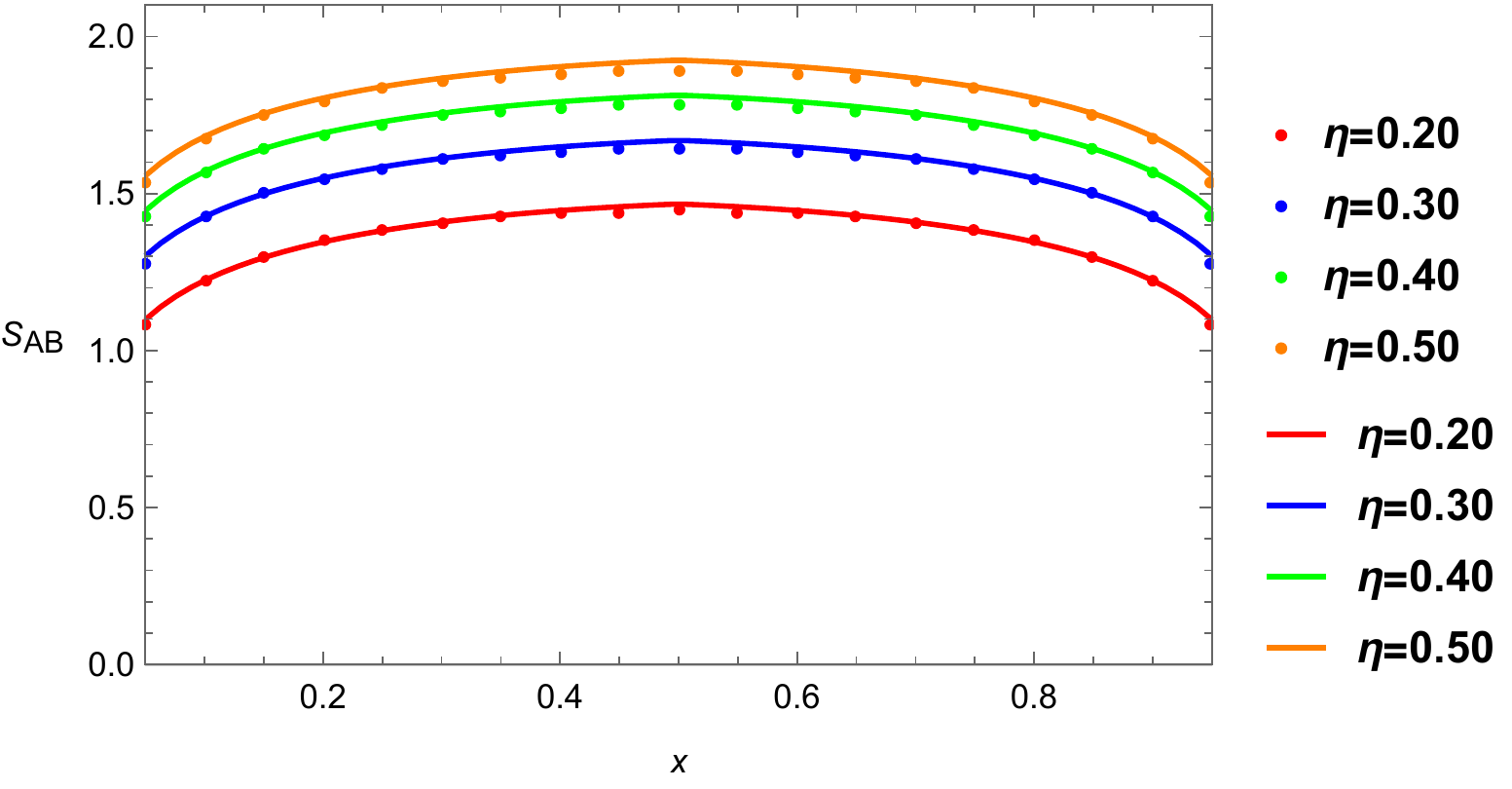}
\includegraphics[width=0.48\textwidth]{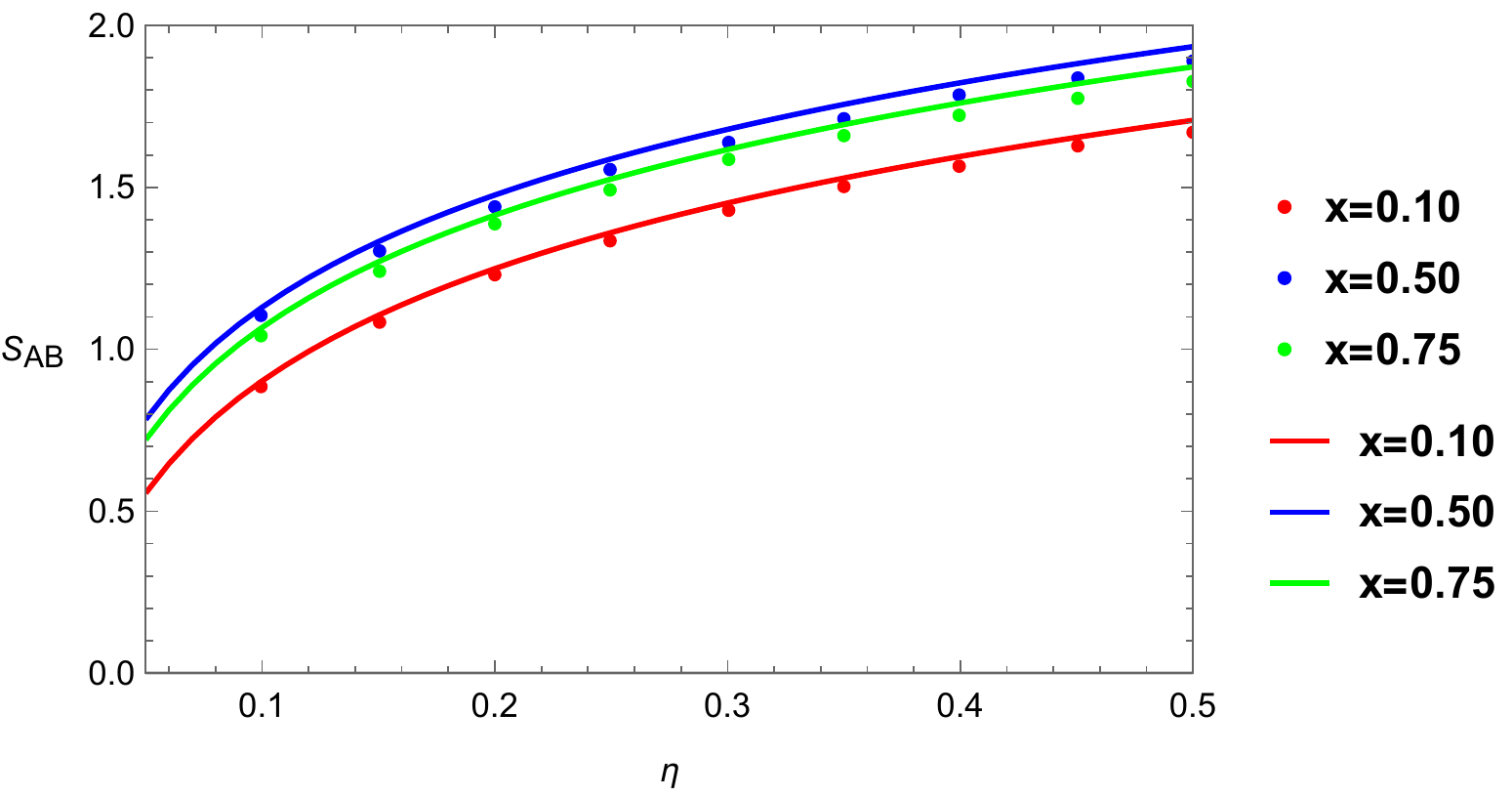}
\caption{\label{decom} Numerical studies of the von Neumann entropy $S_{AB}$ for two intervals in the decompactification limit. Solid continuous curves are plotted according to~\er{Ap}. Each point is calculated by our numerical method up to $k_{\max}=250$. Left: We fix the values of $\eta$ and plot $S_{AB}$ as function of $x$. Right: We fix the values of $x$ and plot $S_{AB}$ as function of $\eta$. We have chosen the cutoff to be $\epsilon^2=0.1$ in all cases.}
\end{figure}

For the decompactification regime, our numerical result will be tested against the von Neumann entropy approximated by the following expansion~\cite{Calabrese:2009ez}
\be \la{Ap}
S_{A \cup B}(\eta \ll 1) \simeq S_{AB}^{W}+\frac{1}{2} \ln \eta - \frac{D'_1(x)+D'_1(1-x)}{2} +\cd,
\ee
where $S_{AB}^{W}$ is the von Neumann entropy computed from~\er{ge} without $\mathcal{F}_{n}(x, \eta)$, and 
\bea
D'_{1}(x)=-\int_{-i \infty}^{i \infty}\frac{dz}{i}\frac{\pi z}{\sin^2 (\pi z)} \ln F_{z}(x).
\eea
We have studied various cases; see Figure~\ref{decom}. For example, with $x=0.25$, $\eta=0.295$, and $\epsilon^2=0.1$ (where it is best approximated~\cite{Calabrese:2009ez, Furukawa:2008uk}), we get $S_{A \cup B} \approx 1.584$ by summing up to $k_{\max}=400$, which is very close to the analytical answer $S_{A \cup B} \approx 1.608$ approximated by~\er{Ap}. The sum has a stable power $p \approx 1.695$, with around $10^{-3}$ relative error.

Note that even with the same choices of numerical constants, the small $x$ expansion would not in general agree with the decompactification limit, which was already indicated in~\cite{Calabrese:2010he}. This is expected as we are not in the exact limit of either $x \to 0$ or $\eta \to 0$.

\begin{figure} 
\centering
\includegraphics[width=0.48\textwidth]{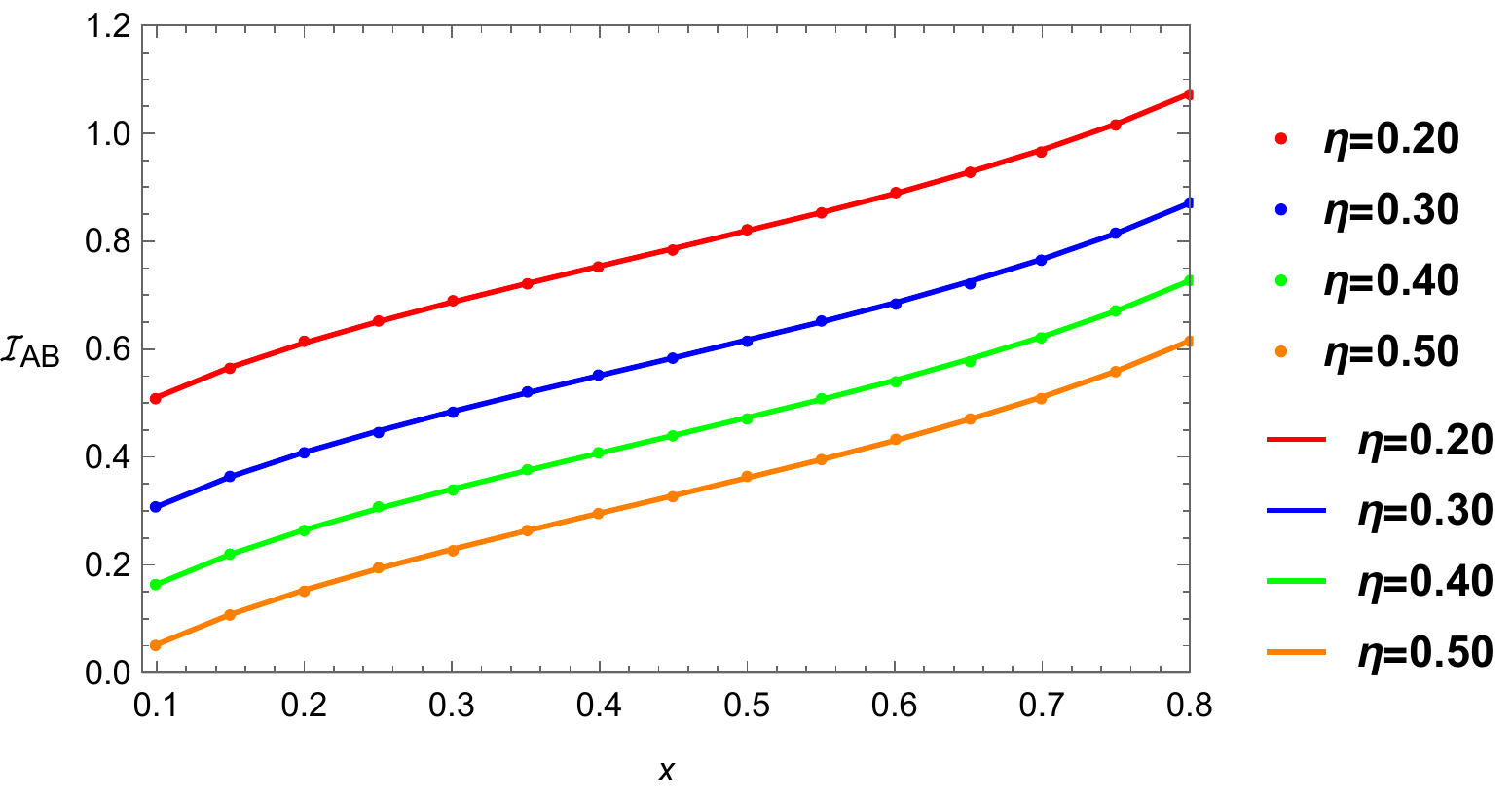}
\includegraphics[width=0.48\textwidth]{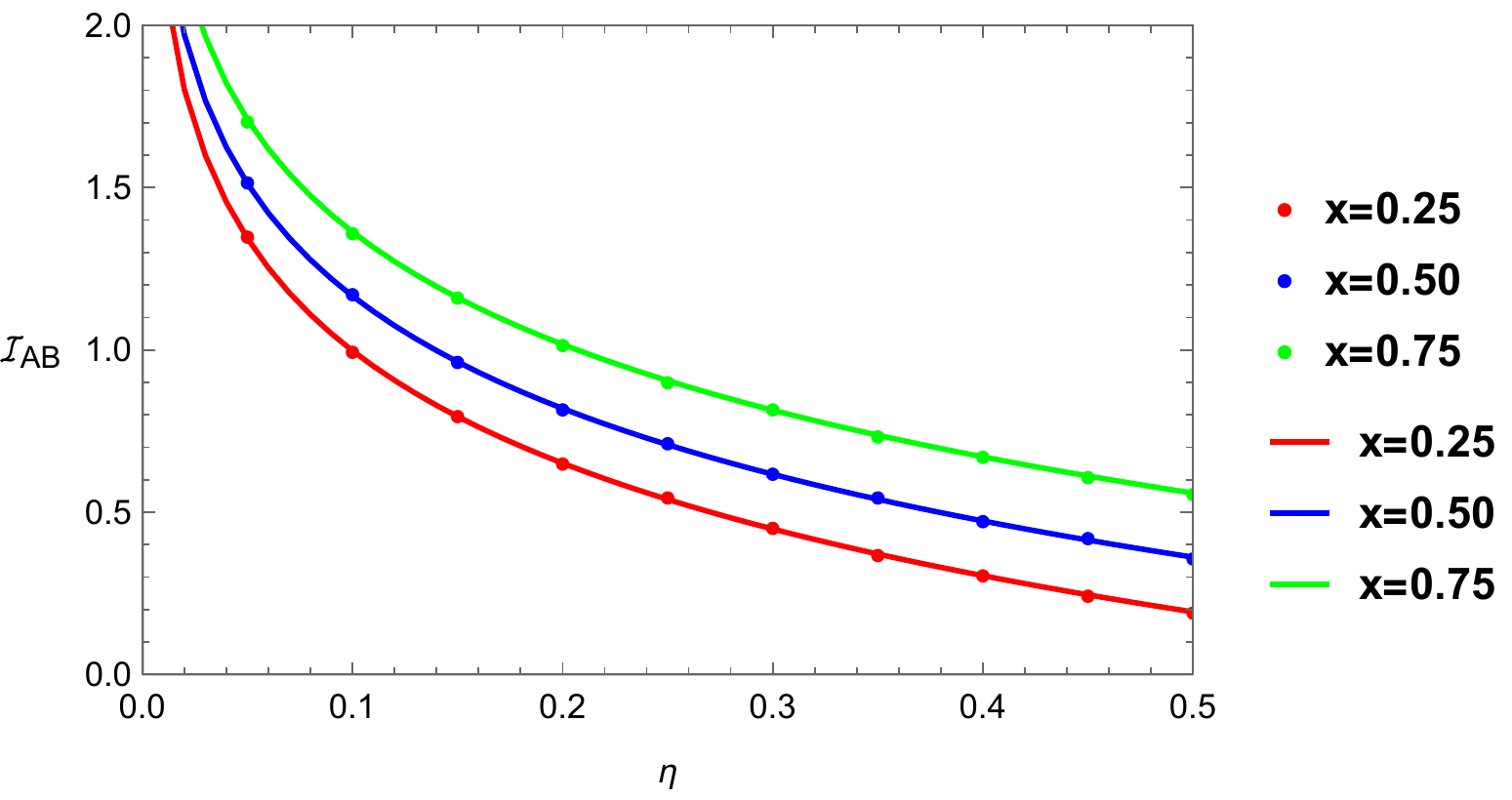}
\caption{\label{demutual} Numerical studies of the mutual information $I_{AB}$ for two intervals in the decompactification limit. Solid continuous curves are plotted according to~\er{mu}. Each point is calculated by our numerical method up to $k_{\max}=200$, with the cutoff set as $\epsilon^2=1$. Left: We fix the values of $\eta$ and plot $I_{AB}$ as function of $x$. Right: We fix the values of $x$ and plot $I_{AB}$ as function of $\eta$.}
\end{figure}

Finally, we may also study the mutual information in the decompactification limit, where we consider the following generating function
\be \la{mut}
f(n-1)= \left (\frac{\epsilon^2}{x_{12}x_{34}} \right )^{\frac{1}{6}(n-\frac{1}{n})}
\left ( \frac{\epsilon^2 x_{13}x_{24}}{x_{12}x_{34}x_{14}x_{23}} \right )^{- \frac{1}{6}(n-\frac{1}{n})} \frac{1}{\mathcal{F}_{n}(x, \eta)}-1
\ee
that according to~\er{mui} should give the mutual information
\be \la{mutt}
I_{AB} = S_A +S_B -S_{AB}.
\ee

The mutual information is approximated in~\cite{Calabrese:2009ez} by
\be \la{mu}
I_{AB} (\eta \ll 1) \simeq I^W_{AB}-\frac{1}{2} \ln \eta + \frac{D'_1(x)+D'_1(1-x)}{2},
\ee
where again $I^{W}_{AB}$ is the mutual information computed from~\er{mut} without $\mathcal{F}_{n}(x, \eta)$. See Figure~\ref{demutual} for our numerical results.  As an example, if we take $x=0.25$, $\eta=0.295$, and $\epsilon^2=1$, we find numerically $I_{AB} \approx 0.456$ by summing up to $k_{\max}=400$, within $10^{-5}$ of~\er{mutt} as well as the answer approximated by~\er{mu}. The sum exhibits a stable power $p \approx 2.457$, within 1\% relative error.

\subsection{Two intervals at finite cross ratio and compactification radius}

For the most general case of finite $x$ and $\eta$, one need to evaluate directly the Riemann-Siegel theta function in~\er{ge} and~\er{RS} numerically.  In this case, an analytical expression for the von Neumann entropy is not yet known. 

We start with~\er{RS} which we reproduce here for convenience
\be\la{RS2}
\mathcal{F}_{n}(x, \eta)=\frac{\Theta(0 | \eta \Gamma) \Theta (0 | \Gamma/ \eta)}{[\Theta (0 | \Gamma)]^2}.
\ee
The expression has a symmetry under $\eta \leftrightarrow 1/\eta $.

To obtain more accurate results in our numerical procedure, one would need to evaluate higher dimensional matrices within the Riemann-Siegel theta function for specific choices of $x$ and $\eta$. This is difficult\footnote{For efforts on computing efficiently the higher dimensional Riemann-Siegel theta function, see~\cite{deconinck2002computing, frauendiener2017efficient}.} to deal with numerically for $k_{\max} \gg 1$. Our approach is to use the identity
\be
\Theta(0 | \eta \tilde{\Gamma})=\eta^{-\frac{n-1}{2}}\bigg(\prod^{n-1}_{k=0} \beta_{k/n} \bigg)^{\frac{1}{2}} \Theta (0 | \Gamma/ \eta),
\ee
to rewrite~\er{RS2} in the following way~\cite{Calabrese:2009ez}:
\be
\mathcal{F}_{n}(x, \eta)= \eta^{\frac{n-1}{2}} \frac{\Theta(0 | \eta \Gamma) \Theta (0 | \eta \tilde{\Gamma})}{\prod^{n-1}_{k=1} F_{k/n} (x) F_{k/n} (1-x)}.
\ee
We use this formula to perform the numerical calculation.  For example, with $x=0.25$, $\eta=0.295$, and $\epsilon^2=1$, summing up to $k_{\max}=15$ we get $S_{A \cup B} \approx 0.747$.

\section{Numerical studies of one interval at finite temperature and length} \la{sec5}

Our next nontrivial example is a single interval at finite temperature and finite length. This example was studied for a 2D free Dirac fermion on a circle using bosonization in~\cite{Azeyanagi:2007bj}. For such a finite system we would need to consider periodic boundary conditions for both space and imaginary time, corresponding to finite size and finite temperature, respectively.  Setting the spatial size to $1$, the two dimensional Euclidean theory thus lives on a torus defined by $z \sim z+1$ and $z \sim z+\tau$, with $\tau=i \beta$ for temperature $\beta^{-1}$. We use $\ell$ to denote the length of the interval.  Using $\Tr(\rho^n)$ calculated in~\cite{Azeyanagi:2007bj}, we find that the generating function is given by
\be \la{TL}
f(n-1)= \prod_{k=-\frac{n-1}{2}}^{\frac{n-1}{2}} \bigg|\frac{2 \pi \epsilon \eta(\tau)^3}{\theta_{1}(\ell | \tau)} \bigg|^{\frac{2 k^2}{n^2}} \frac{|\theta_{\nu}(\frac{k \ell}{n}| \tau)|^2}{|\theta_{\nu}(0|\tau)|^2}-1,
\ee
where $\epsilon$ is the UV cutoff\footnote{Note that one should keep at least $\epsilon \leq \beta$ for the procedure to be physical. Otherwise, we would be probing states beyond the UV cutoff.} and $\nu$ is determined by the boundary condition for the fermion.   In particular, we will study the case of $\nu=3$ that corresponds to the Neveu-Schwarz (NS-NS) sector. Note that $\eta(\tau)$ is the Dedekind eta function defined as
\be
\eta(\tau) \equiv q^{\frac{1}{24}} \prod_{n=1}^{\infty}(1-q^n),
\ee
where $q=e^{2 \pi i \tau}$. The Jacobi theta functions $\theta_{1}$ and $\theta_{3}$ are defined as
\be
\theta_{1} (z|\tau) \equiv \sum_{n=-\infty}^{n=\infty} (-1)^{n-\frac{1}{2}} e^{(n+\frac{1}{2})^2 i \pi \tau} e^{(2n+1) \pi i z} \,,\qqu
\theta_{3} (z|\tau) \equiv \sum_{n=-\infty}^{n=\infty} e^{n^2 i \pi \tau} e^{2 n \pi i z} \,.
\ee

The exact expressions for the von Neumann entropy are only known in high-temperature and low-temperature expansions.
For the high-temperature expansion, the von Neumann entropy is
\bm \la{hiT}
S^{H}_{A} =   \frac{1}{3} \ln \bigg(\frac{\beta}{\pi \epsilon} \sinh{\frac{\pi \ell}{\beta}} \bigg)+\frac{1}{3} \sum_{m=1}^{\infty} \ln \frac{(1-e^{2 \pi \frac{\ell}{\beta}}e^{-2 \pi \frac{m}{\beta}})(1-e^{-2 \pi \frac{\ell}{\beta}}e^{-2 \pi \frac{m}{\beta}})}{(1-e^{-2 \pi \frac{m}{\beta}})^2}  \\
\quad +2 \sum_{l=1}^{\infty} \frac{(-1)^l}{l} \bigg( \frac{\frac{\pi \ell l}{\beta} \coth{\frac{\pi \ell l}{\beta}}-1}{\sinh{\frac{\pi l}{\beta}}} \bigg).
\em
Note that the first term reproduces the infinite length, finite temperature von Neumann entropy~\cite{Holzhey:1994we, Calabrese:2004eu, Calabrese:2009qy}
\be \la{te}
S^{H}_{A}=\frac{1}{3} \ln \bigg( \frac{\beta}{\pi \epsilon} \sinh{\frac{\pi \ell}{\beta}} \bigg),
\ee
which is universal. Numerically, we take the choices of $\beta =0.9$, $\ell=0.5$, and $\epsilon=0.1$. By summing up to $k_{\max}=700$ for~\er{TL}, we get $S_{A} \approx 0.580$, which is within $10^{-3}$ of the analytical answer $S_{A} \approx 0.582$ from~\er{hiT}, with a stable power $p \approx 1.873$, within $10^{-2}$ relative error. 

\begin{figure} 
\centering
\includegraphics[width=0.7\textwidth]{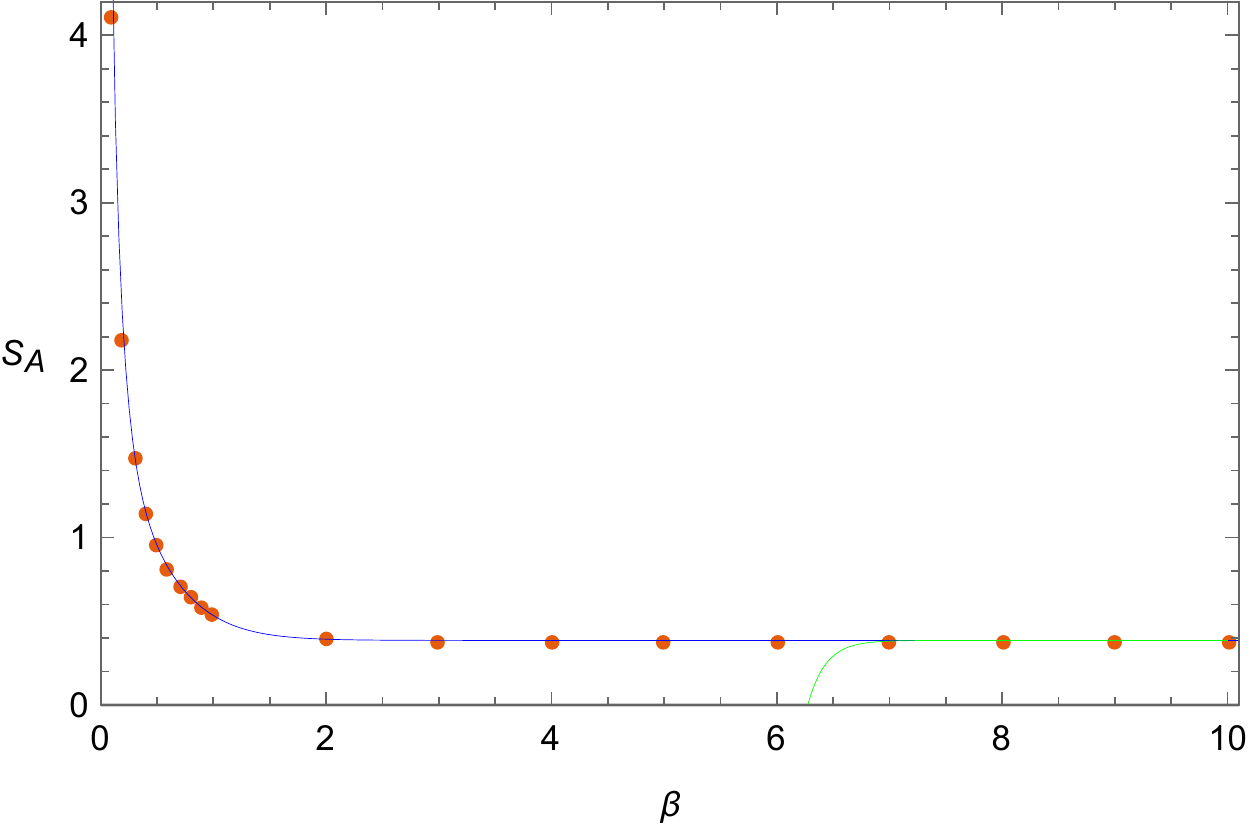}
\caption{\label{inter} Numerical studies of a single interval $A$ at finite temperature and length. We fix the interval length $\ell=0.5$ and the cutoff $\epsilon=0.1$, and plot $S_A$ as a function of $\beta$. The blue curve is plotted by the high-temperature expansion~\er{hiT}. The green curve is plotted by the low-temperature expansion~\er{loT}. Each point is calculated by our numerical procedure up to $k_{\max}=200$. One can see that the high-temperature expansion fits well in the plot, but the deviation with our numerical method increases slowly for larger $\beta$ not shown here.}
\end{figure}

For the low-temperature expansion, the von Neumann entropy is
\bm \la{loT}
S^{L}_{A} =   \frac{1}{3} \ln \bigg(\frac{1}{\pi \epsilon} \sin \pi \ell \bigg)+\frac{1}{3} \sum_{m=1}^{\infty} \ln \frac{(1-e^{2 \pi i \ell}e^{-2 \pi \beta m})(1-e^{-2 \pi i \ell}e^{-2 \pi \beta m})}{(1-e^{-2 \pi \beta m})^2}  \\
\quad +2 \sum_{l=1}^{\infty} \frac{(-1)^{l-1}}{l} \bigg( \frac{1-\pi l \ell \cot{\pi \ell l}}{\sinh{\pi l \beta}} \bigg).
\em
Similarly, we see that the first term reproduces the finite size, zero temperature von Neumann entropy~\cite{Holzhey:1994we, Calabrese:2004eu, Calabrese:2009qy}
\be \la{le}
S^L_{A}=\frac{1}{3} \ln \bigg(\frac{1}{\pi \epsilon} \sin{\pi \ell} \bigg),
\ee
which again is universal. Numerically, we take the choices of $\beta =10$, $\ell=0.5$, and $\epsilon=0.1$.  By summing up to $k_{\max}=700$ for~\er{TL}, we get $S_{A} \approx 0.385$, which is within $10^{-3}$ of the analytic result $S_{A} \approx 0.386$ from~\er{loT}, with a stable power $p \approx 1.922$, within $10^{-3}$ relative error. 

Even though the analytic expressions are known only for the high and low $\beta$ regimes, we can numerically interpolate between the two regimes using the generating function (see Figure~\ref{inter}).

\section{Power-law convergence of the generating function} \la{sec6}

The various field theory examples that we have considered so far all exhibit power-law convergence at $w=1$ of the series~\er{hws}.  In other words, we have
\be \label{pow}
\frac{\tilde{f}(k)}{k} \sim k^{-p}
\ee
at large $k$ with a power $p$ that depends on each specific example. This leads to a natural question: what type of eigenvalue distributions for a density matrix $\rho$ would exhibit such power-law behaviors? 

To answer this question, we define the eigenvalue distribution $P(x)$ so that the number of eigenvalues within a small range $[x,x+dx]$ is $P(x)dx$.  We may rewrite $\Tr(\r^n)$ in terms of $P(x)$:
\be\la{trnp}
\Tr(\r^n) = \int_0^1 dx P(x) x^n.
\ee
In particular, $\Tr \r=1$ means
\be\la{trp}
\int_0^1 dx P(x) x =1.
\ee
In the case of a finite-dimensional density matrix with eigenvalues $\{\r_i\}$, we have $P(x) = \sum_i \d(x-\r_i)$.

Using Eq.~\er{trnp}, we may write Eq.~\er{tfk} as
\be\la{tfkp}
\td f(k) = \int_0^1 dx P(x) x (1-x)^k.
\ee
We would like to find the behavior of this integral for large $k$.  We expect the integral to be dominated by small $x$ when $k$ is large, due to the presence of $(1-x)^k$.

Let us therefore consider a power-law ansatz for the behavior of the eigenvalue distribution $P(x)$ near zero eigenvalue:
\be
P(x) \sim x^\g,\qqu
\text{as} \qu x \to 0.
\ee
Convergence of the integral in Eq.~\er{trp} then requires $\g>-2$.  Substituting this into Eq.~\er{tfkp}, we find at large $k$
\be
\td f(k) \sim \int_0^1 dx x^{\g+1} (1-x)^k = \fr{\G(k+1)\G(\g+2)}{\G(k+\g+3)} = \fr{1}{k^{\g+2}} \[\G(\g+2)+\cO(k^{-1})\].
\ee
This is indeed a power law for large $k$.  Comparing it with our definition of the power $p$ in Eq.~\er{pow}, we find
\be\la{pg}
p = \g+3.
\ee
As we mentioned previously, Eq.~\er{trp} requires $\g>-2$, giving $p>1$, which is precisely the necessary and sufficient condition for the sum~\er{hws} to converge at $w=1$.

Eq.~\er{pg} describes how the power-law behavior of the sum~\er{hws} is related to the small eigenvalue behavior of the eigenvalue distribution $P(x)$.  Therefore, using the power $p$ that we have determined numerically in the field theory examples studied in previous sections, we may now predict that their eigenvalue distribution $P(x)$ must scale like $x^{p-3}$ for small eigenvalue $x$.

As we alluded to briefly in section~\re{sec3}, this power-law convergence is closely related to the infinite-dimensional Hilbert space of quantum field theories. Let us use $\{\r_i\}$ to denote the eigenvalues of the density matrix.  The generating function~\er{gdf} becomes
\be\la{gri}
G(z;\rho) =-\sum_i \bigg[ \r_{i} \ln \frac{1-\r_i z}{1- z} \bigg].
\ee
It has branch cuts along $z \in [1, 1/\r_{\min}]$ where $\r_{\min}$ is the smallest eigenvalue of the density matrix $\rho$. After the M\"obius transformation to $w$, the branch cuts are along $w \in [\frac{1}{1-\r_{\min}}, \infty)$. A necessary condition for the series~\er{hws} to be power-law convergent at $w=1$ is that its radius of convergence must be 1.  This means $\frac{1}{1-\r_{\min}} = 1$, or $\r_{\min} = 0$.\footnote{It is manifest from Eq.~\er{gri} that in order to have power-law convergence, it does not help for some of the eigenvalues to be precisely zero; instead, there must be an accumulation of eigenvalues near zero.} This is true for density matrices in continuous quantum field theories with infinite-dimensional Hilbert spaces.

\section{Discussion} \la{secdis}

In this paper we have shown that the von Neumann entropy may be obtained by assembling the traces $\Tr (\rho^n)$ for all positive integers $n$ into a generating function of an auxiliary complex parameter $z$, analytically continuing in $z$, and then taking the limit as $z \to - \infty$. Our construction demonstrates that the analytic continuation in $z$ exists when $\rho$ is a density matrix. We showed how the procedure may be carried out analytically for the cases of one interval and of two colinear intervals in certain limits, and we demonstrated that a simple variant of the method also leads to numerical evaluations of the von Neumann entropy in a practical, reliable way.

Many open questions and future directions for investigation remain. Most of the cases that we studied were selected because they are simple enough to exhibit the method clearly, but complicated enough so that the standard replica ``analytic continuation'' in $n$ is not automatic.  Firstly, it is urgent to see whether our method can analytically produce the von Neumann entropy in cases that have eluded direct analytic continuation in $n$.  One such case is the example of two intervals at finite cross ratio for a free boson with a general compactification radius.  Another interesting example involves the contributions of replica non-symmetric saddle point solutions in holography, which are difficult to analytically continue in $n$ but have been shown in~\cite{Dong:2020iod} to give important enhanced corrections to the Ryu-Takayanagi formula~\cite{Ryu:2006bv,Lewkowycz:2013nqa} at holographic entanglement phase transitions.

Secondly, our method appears to have a broader regime of applicability than what might be expected from our derivation -- in particular, the method often applies to parts of $\Tr (\rho^n)$ with ``easy factors'' stripped off (as in footnote~\re{ft1}) and we also conjecture that it applies to the mutual information (as in section~\re{sec3}).  It would be very interesting to understand precisely how broadly our method applies and why.

Finally, an alternative starting point for the replica method, which recently appeared in~\cite{Witten_2019}, is from the functions $\Tr (\rho^{1/n})$ instead of $\Tr (\rho^n)$. It will be interesting to see whether our method can be adapted to handle such cases as well.

\section*{Acknowledgments}

It is a pleasure to thank Per Kraus, Don Marolf, and Edward Witten for useful conversations. The research of ED is supported in part by the National Science Foundation under grants PHY-16-19926 and PHY-19-14412. XD and CW are supported in part by the National Science Foundation under Grant No.\ PHY-1820908 and by funds from the University of California.  XD is grateful to the Institute for Advanced Study and the Kavli Institute for Theoretical Physics (KITP) where part of this work was developed.  The KITP was supported in part by the National Science Foundation under Grant No.\ PHY-1748958.

%%%%%%%%%%%%%%%%%%%%%%%%%%%%%%%%%%%%%%%%%%%%%%%%%%
\bibliographystyle{JHEP}
\bibliography{bibliography}

\end{document}